# Forecasting COVID 19 growth in India using Susceptible-Infected-Recovered (S.I.R) model

Jay Naresh Dhanwant and V. Ramanathan*

Department of Chemistry, IIT(BHU) Varanasi, Uttar Pradesh 221005,India
* Author to whom correspondence to be addressed at vraman.chy@iitbhu.ac.in

## Abstract
This work covers the analysis of the COVID 19 spread in different countries, and dealing the main feature of COVID 19 growth, that is the spread due to social-contact structure, which is governed by the parameter β. The dependency of this parameter β on the transmission level in a society gives a sense of effectiveness of the measures taken for social distancing. A separate algorithm is hardcoded in python using Scipy which learns the social-contact structure and gives a suitable value for β, which has the major impact on the outcome of the result. Forecasting for the epidemic spread in India was done, and it was found that the strictness at which social distancing in India is done, is insufficient for the growth of COVID 19.

## Introduction

The novel coronavirus disease (COVID-19) had its outbreak initially in Wuhan, China. It is an infectious disease caused by a new kind of virus. The disease causes respiratory illness, with symptoms such as cough, fever and difficulty in breathing in several cases. This virus spreads primarily through contact with the virus that gets carried by the body fluid of an infected person mainly when they cough or sneeze. With the rapid spread of this virus, World Health Organization has declared it as a pandemic. At this time, there is no specific vaccine or treatment. This virus has infected more than 8,50,000 people and taking more than 40,000 lives [1]. All the countries across the world are trying to halt the virus. The only accepted way to attenuate the growth is to practice social distancing. Major steps are taken by various governments across the globe, and the most critical steps are imposing lockdowns on countries. With varied implementation of lockdown, there have been some outbreaks, including in Italy and Spain and this work consists of the analysis and prediction of such measures. Similar measures have been taken in India to minimize social contacts, including the shut-down of various organizations, lockdown of schools and colleges, Junta Curfew, and a 21-day lockdown. Since these measures have immense pressure on economy and is important for containing the Coronavirus, quantitative estimates are imperative to learn the impact of spread which will help in planning policies. Given the paucity of such quantitative estimates, the predictions given in this paper becomes critical and to know when the changes are required. This paper consists of evaluating these metrics and thereby come up with quantitative estimates using customized Susceptible-Infected -Recovered (SIR)

Previous work for prediction of the spread of COVID 19 has been done using many different machines learning algorithms, including neural networks for deep learning, polynomial fitting, exponential smoothing and ARIMA [2-8]. Neural Networks end up overfitting the data, and polynomial fitting using a third-degree polynomial is also found to be overfitting with a very high bias [3]. This is because the trend in the epidemic spread assumes different nature in different phases overtime. A polynomial fit will give an appropriate result for an instance, but as soon as the society relaxes lockdown or changes social-contact pattern in any way, these model cause a lot of deviations. A more appropriate method is found out to be an analytical solution to the differential system of equations using a three-compartment system SIR model described below.

This model divides the entire population into three categories, Susceptible, Infected and Recovered. Each compartment is expected to have the same characteristics.

☐ S: Susceptible Population

☐ I: Infectious Population

☐ R: Recovered Population

The three categories are interrelated with each other with the parameters β and γ. $dS/dt = -\beta*I*S$, $dI/dt = \beta*I*S - \gamma*I$, $dR/dt = \gamma*I$. The main drawback of this model is the presumption of the parameter β [4,5].

hardcoded in python using Scipy which learns the social-contact structure and gives a suitable value for β, which has the major impact on the outcome of the result. Forecasting for the epidemic spread in India was done, and it was found that the strictness at which social distancing in India is done, is insufficient for the growth of COVID 19.

In the later section, the model has been applied to Hubei-China, and India for generation of hypothesis. This hypothesis was then compared to actual data to show the versatility of the model in exponential growth phase and a decline phase as well. Finally the metric for comparison of the lockdown was done and a detail data analysis of the death rate and its impact on India is discussed.

## Mathematical Model

Rate of change of Susceptible Population = $\frac{dS}{dt} = -\beta * I * S$

Rate of change of Infectious Population = $\frac{dI}{dt} = \beta * I * S - \gamma * I$

Rate of change of Susceptible Population = $\frac{dR}{dt} = \gamma * I$

This model was implemented in python using scipy for solving the differential equations [6]. Key influencers here are the parameters β and γ: which are described as:

- β: Decides how much the disease can be transmitted due to exposure. An interesting point here is, this β can be different for a same kind of virus in different society. A society with less socialization will have a lower value of β.
- γ: is a parameter expressing how much the disease can be recovered in a specific period
- α and ⁻α: These are hidden parameters that are not explicit in the above equations. This is the measure of symptomatic cases + asymptomatic cases

To take α into account, total number of cases were simulated. The ratio between actual number of infections and the confirmed reported cases is found to be 4.5 from data analysis [5], it decreases over time if the exponentiality in the growth decreases
The entire analysis revolves around β as there is no medical treatment found yet for the virus, and the only way to control it is by reducing transmission which this parameter governs.

## Results and Discussion

### Factors influencing the infectious population:

Initially, the entire population is susceptible to the infection, and very negligible amount of people bring this infection into the country. As the disease spreads, infectious section starts increasing, along with the number of recovered cases. The rate at which a susceptible population funnels into infected population depends on various factors, namely:

I. How fast the disease spreads naturally (unique to the virus)
II. Social Distancing Measures of the country (unique to the society)
III. How rapid the tests are done for detection? (unique to medical infra structure)

All these factors contribute to the rate at which Susceptible population is converted into infectious population, this parameter is termed as β

- Rate at which the people are recovered: This is mostly unique to the virus; this parameter is termed as γ in the model
- A hidden parameter α: There is a period between which a person gets infected, and is confirmed positive. Hence the total number of infected people is greater than the total number of confirmed positives in an exponential growth phase, and the total number of infected people is close to total number of confirmed cases when the growth is halted. This is the reason why many developed countries failed to halt the growth of the virus in exponential phase. The influence of α is shown in Figure 1 for Hubei and Italy respectively.

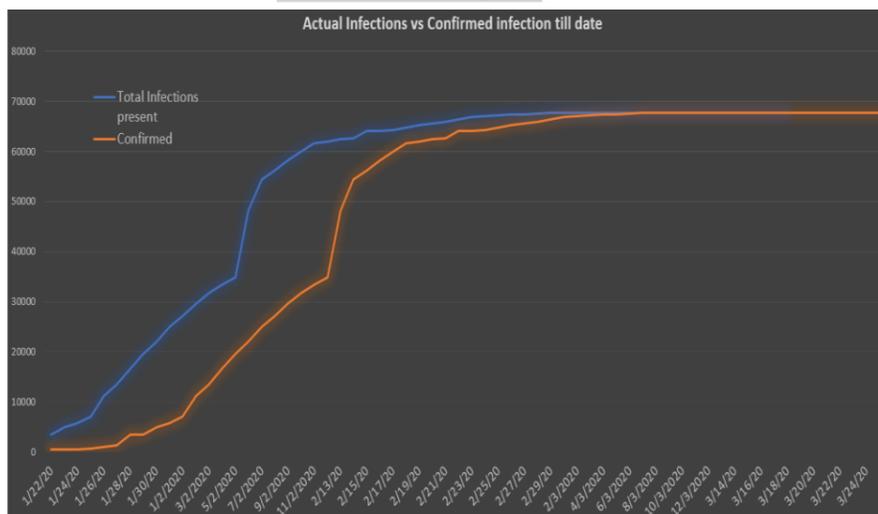 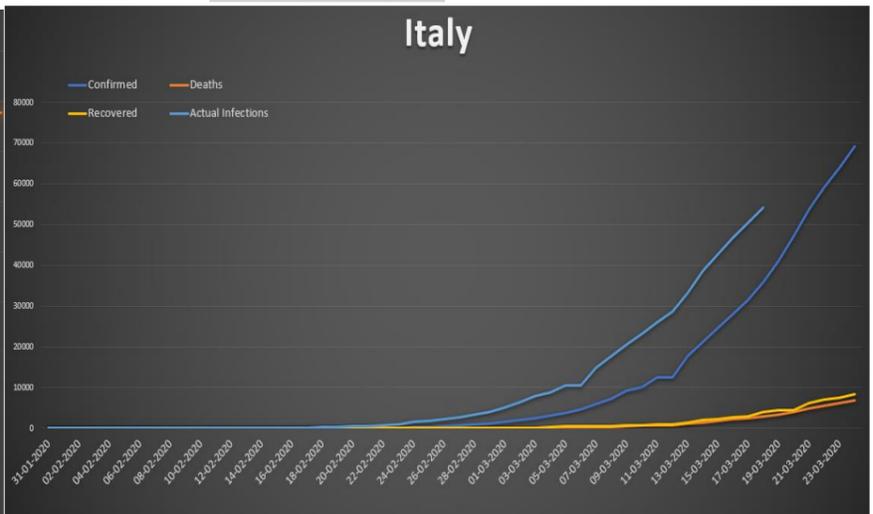

*Figure 1: Influence of α: Ratio between the confirmed equations and the total number of infections is 0.2 initially and goes close to one as the epidemic is halted*

It can be inferred from figure 1 that there is an increase in the reported confirmed cases as well as the actual cases in Hubei, after which the two curves came closer making the scenario controllable. The number of reported cases kept on rising after the halt of increase in actual cases after lockdown and this shows the influence of incubation period. These cases were majorly the ones who were infected before the lockdown in India and their incubation period continued after lockdown. As expected, the two parameter meets when there is no new cases and the pandemic is contained. Failure of controlling the situation, and the number of unreported new cases kept on rising in Italy causing the pandemic to explode.

The algorithm available in established libraries didn't give appropriate results for India, since the condition in which a country goes through changes as the level of measures increases with the severity of the disease.

SIR model was hardcoded for COVID 19 using python and Scipy. A custom loss function was framed and the value of β with minimum loss was taken by applying gradient descent and iterating it over a range of β. While the value of α was adjusted internally for each day, as the exponentially decreases, α comes closer to one. As a proof of principle the developed algorithm was validated with two sets of data.

Initially, an 80:20 split was performed and β was learned from the training data, then the prediction was made for Hubei, China during its lock-down period. Due to small number of data, a simple model turned out to be the best fit, as the data just wasn't enough to draw insights from other possible options like Neural Networks, Deep Learning or ARIMA as seasonality was immaterial in the situation under lockdown conditions. Using the model, the value of β learned, turned out to be 0.1 and this was used for predicting the test data. The result is depicted in Figure 2 which shows a comparison between the actual result and hypothesis. Further proof of principle was carried out for the data available from India during the lock-down period (until March 30, 2020). This result is summarised in Figure 3 where once again the robustness of the developed model is successfully tested through validation by the real data. The two validations clearly suggest that the model works with reasonable accuracy in both the growth as well as the decline phase.

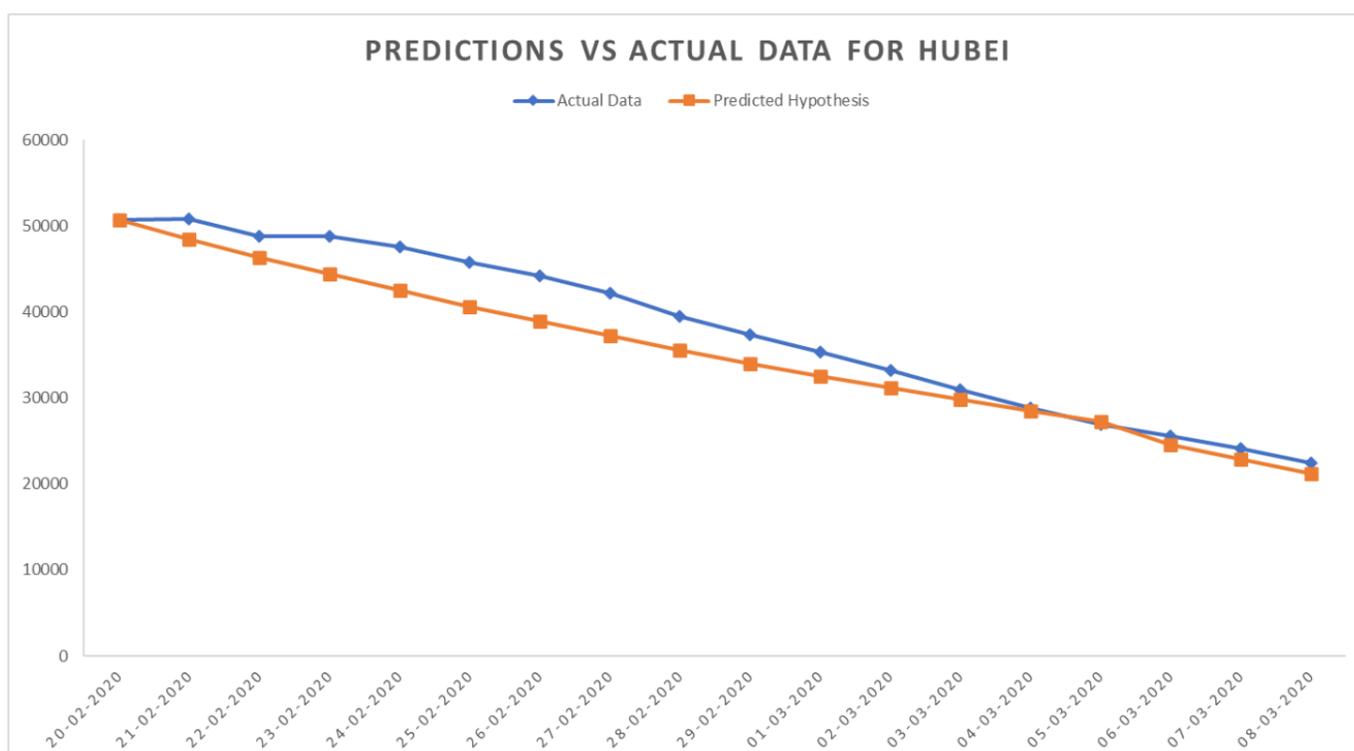

Figure 2: Proof of principle: Validation of the developed model with actual data from the lock-down period in Hubei, China

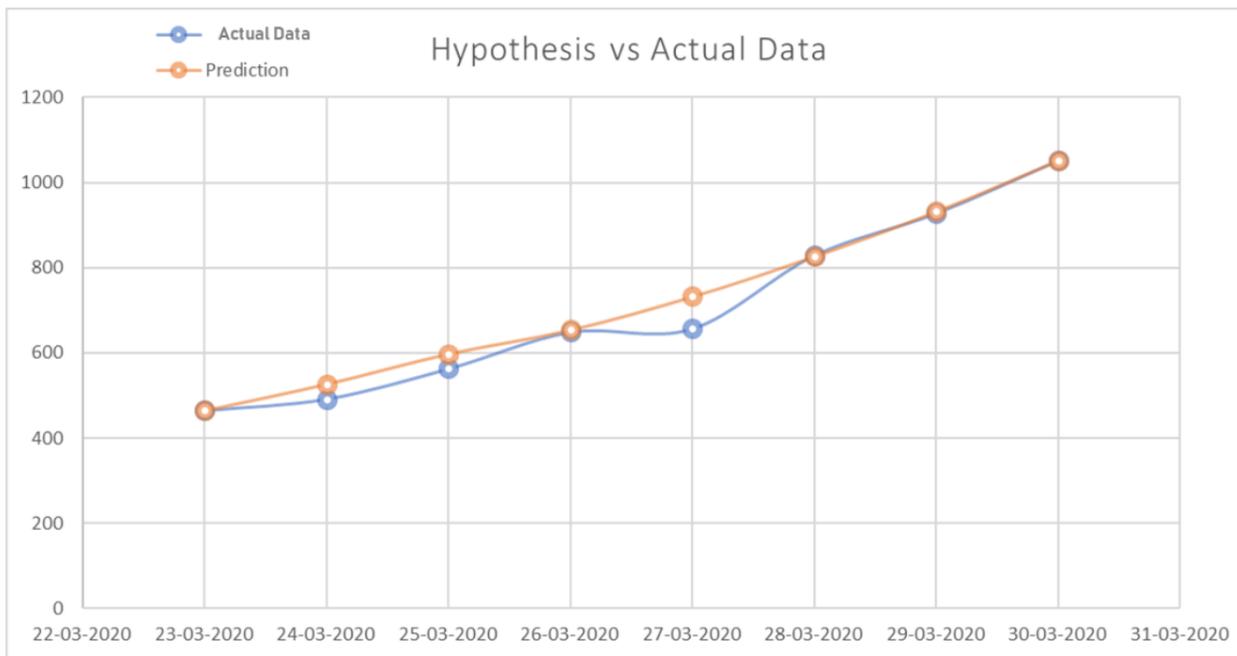

Figure 3: Proof of principle: Validation of the developed model with actual data from the lock-down period in India

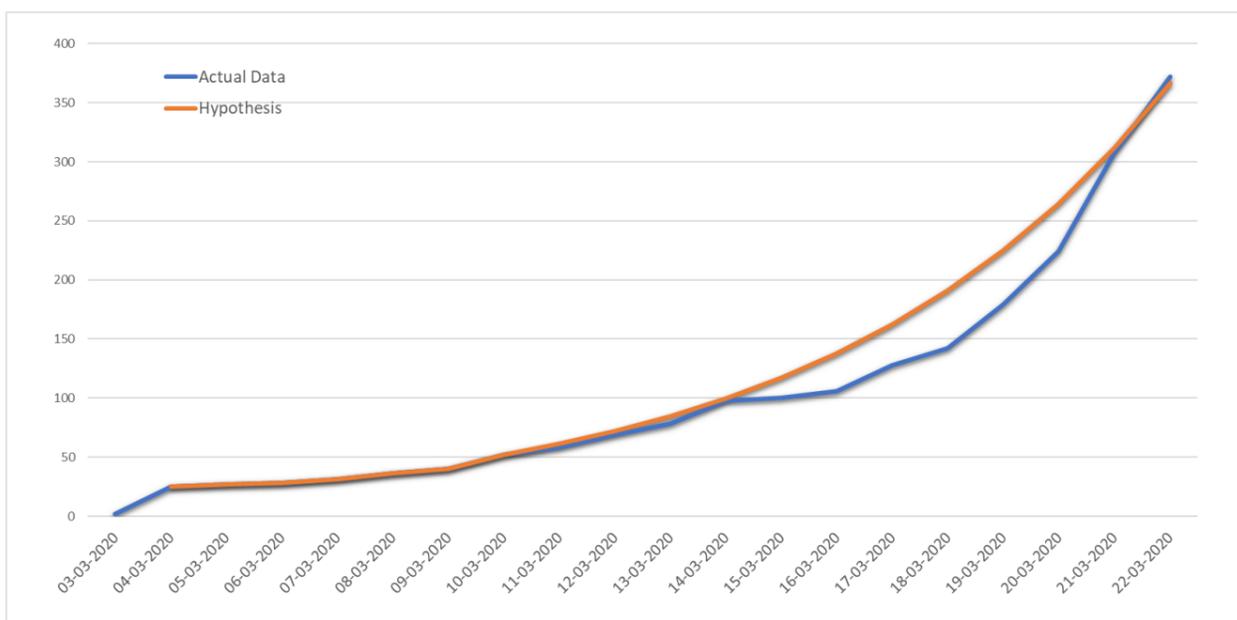

Figure 3: Proof of principle: Validation of the developed model with actual data before the lock-down period in India

Taking further, the two values of parameter β were learned, one when social distancing is compromised, but the basic measures are taken, and another when country imposes a lockdown. This was learned by making a custom loss function which was minimized by using gradient descent on the loss function. β = 0.3536 was observed when the curfew wasn't imposed, and β = 0.2627 was observed under the lockdown conditions. This signifies the lockdown period caused a considerable impact in the spread of corona virus. Forecasting was done to find out whether the lockdown was able to prevent or halt the growth of COVID 19. Using these learned parameters the forecast for India is presented in Figures 4 which spans the entire lock-down period in India. The model predicts that the number of active cases will go from 568 to 6424 in 21 days. Model suggests that, an increase of 947 new confirmed cases is predicted in week one, week 2 with 1604 new cases following 3310 new cases in week 3. It must be highlighted that the predictions made by Singh and Adhikari [4] feature a sharp decrease in the number of cases after 25[th] March but the actual data is quite the opposite which is truly predicted by the model presented here.

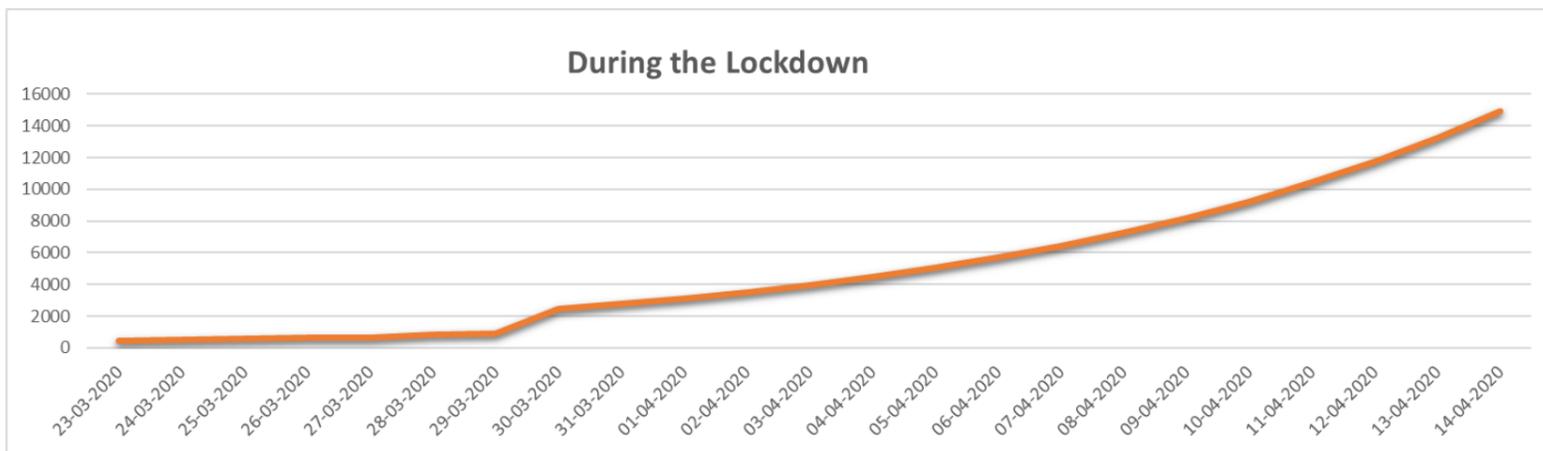

Figure 4: Predictions for India during the entire span of the lock-down period.

Further, using the other value for β that is in the absence of any lock-down, the model developed by us predicts a further increase in the number of cases upto 0.145 million, quite contrary to the predictions of Singh and Adhikari [4]. This is depicted in Figure 5.

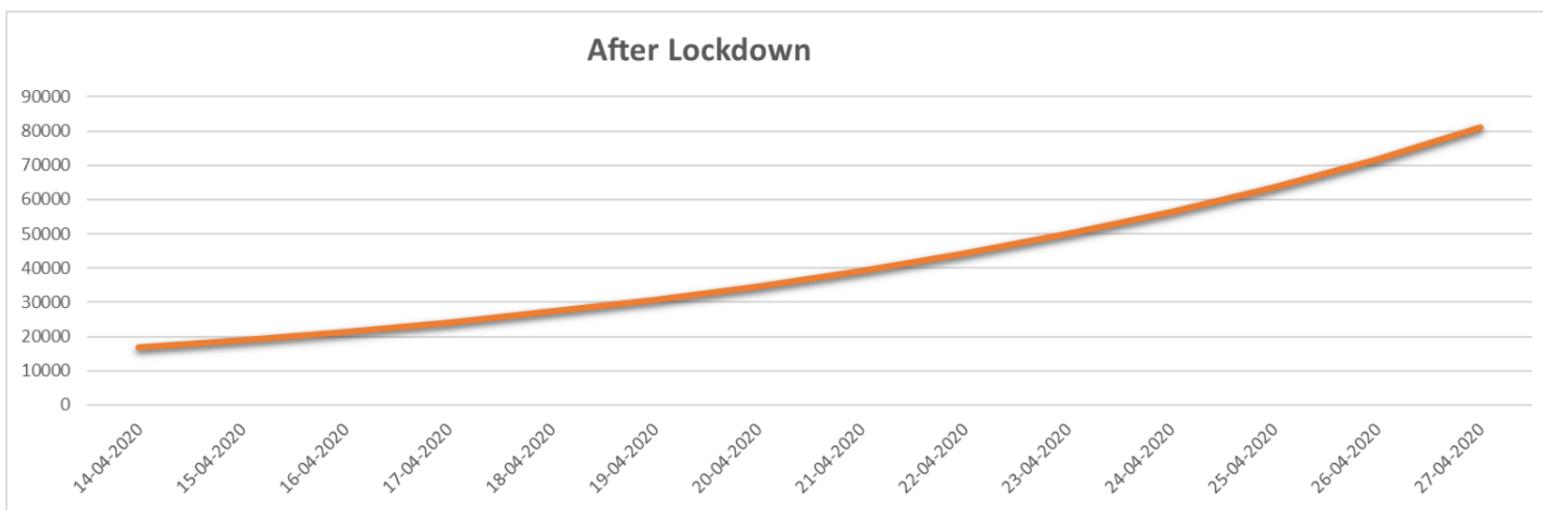

Figure 5: Predictions for India during the entire span of the lock-down period.

## Effectiveness of lockdown in Hubei and India

Social distancing is the only proven solution for halting corona virus growth as of now. Nation practices strict social distancing in the lockdown, but still due to several factors, we see the failures in some lockdowns. Parameter β governs the transmission efficiency of the virus in given society, since lockdowns and social distancing measures focuses on reducing the transmission, reduction in β for different stages will reflect the how better the lockdown was performed.

The β for a complete shutdown in Hubei comes out to be 0.1 on the training data and 0.8 on complete dataset through the period, whereas β in India remains at 0.2627.

As expected of a lockdown, in India too, β is decreased from 0.3536 to 0.2627. But this value is not good enough to halt the growth of corona virus in India.

Similarly, this can be used for other countries as well, to determine the outcome of a lockdown, based on the data generated.

## Death Rate:

This is defined as the number of deaths / confirmed cases. this metric is of importance as the death rate is merely 2% for the virus, but more fatal for elderly people and those having other complications. But this increases as the number of infections goes up, **this can go up to 10%** and in a country having a population of 134 crore, 10% death rate phase has to be avoided at any cost. Figure 6 shows the death rate in Italy over the time, which shoots up as soon as Italy crosses a figure of 12,000 COVID 19 patients. The death rate reaches as high as 10% and is dependent on the percentage of the elderly people and number of beds per 1000 citizens. The number of beds per 1000 people in Italy is 3.18 while it is around 0.55 in India. The total number of beds available in India is around 7 Lakhs, which is one seventh of Italy. The percentage of people over 60 years is 23% in Italy whereas in India, it is around 8%. It is safe to assume a death rate of 8% when the number of patients overwhelms as the mortality of COVID 19 is less about the disease and more about the health care system and mass of patients. Threshold for the spike in death rate which is 12,000 patients in Italy, this is estimated to be 84,000 patients in India, based on the analysis of data, for such spike in death rate, and this spike can very well be between 7% – 10% based on the data provided.

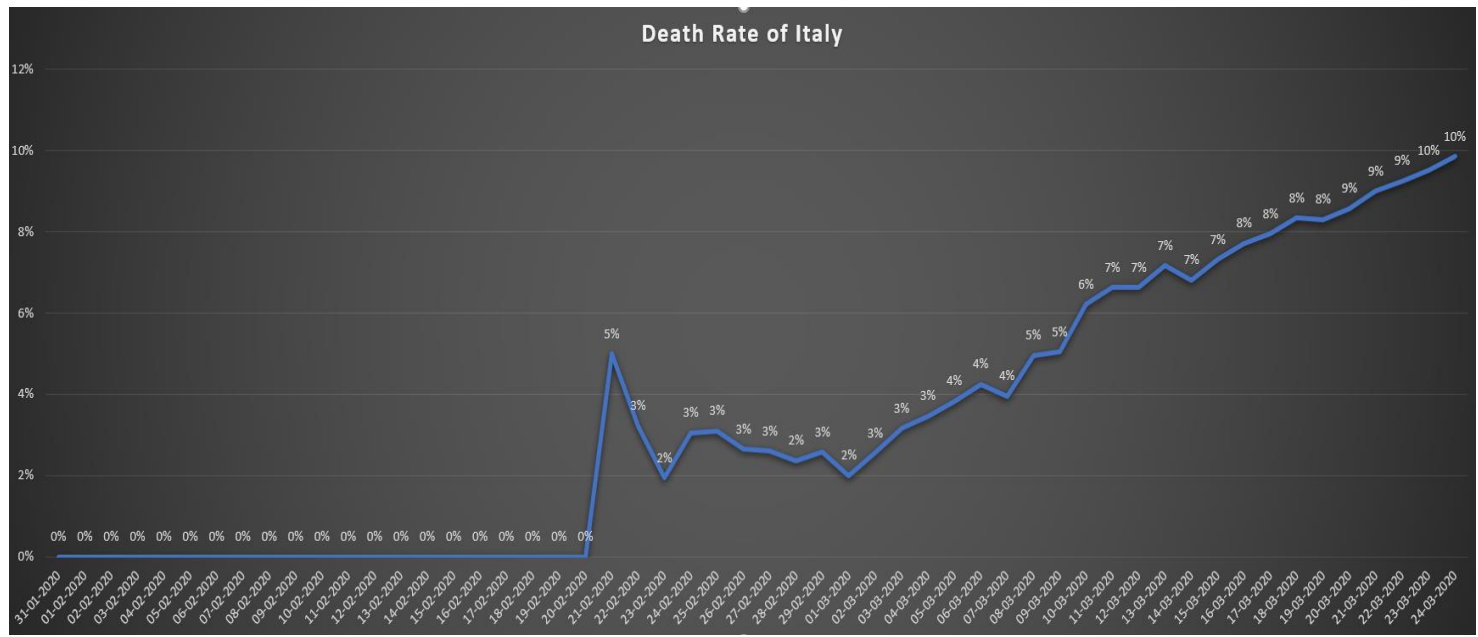

# Summary and conclusion:

- The model developed by us and reported here has a distinct advantage over other models as it gives parameters tuned with the data, that provides insight over the actual situation.
- It accurately fit's exponential growth, linear growth, and a natural decline in the pandemic.
- When a government takes a major step related to Corona Virus, the feature tuning algorithm, discussed above, is trained on the data to get the new changed value of β as a result.
- Even during the lockdown, there will be no decrease in the number of cases as revealed by the data.
- As this model has a high bias, it will get more accurate after training it on some more lock-down data observed in near future
- Actual Number of cases at any day in near future will be much higher than the total number of cases reported. Convergence of these two values will result in the halt of exponential growth.
- Graph suggests that the exponential growth will be slowed down in the lock-down, but we need to observe a stricter social distancing for a halt of growth.